\documentclass[showpacs,preprintnumbers,amsmath,amssymb,nofootinbib]{revtex4}
\usepackage{latexsym,epsf}
\usepackage{graphicx}
\usepackage{bm}
\usepackage{epsfig}
\begin{document}

\newcommand{\za}{\alpha}
\newcommand{\zb}{\beta}
\relax
\newcommand{\TeV}{\,{\rm TeV}}
\newcommand{\GeV}{\,{\rm GeV}}
\newcommand{\MeV}{\,{\rm MeV}}
\newcommand{\keV}{\,{\rm keV}}
\newcommand{\eV}{\,{\rm eV}}
\newcommand{\Tr}{{\rm Tr}\!}
\renewcommand{\arraystretch}{1.2}
\newcommand{\be}{\begin{equation}}
\newcommand{\ee}{\end{equation}}
\newcommand{\bea}{\begin{eqnarray}}
\newcommand{\eea}{\end{eqnarray}}
\newcommand{\ba}{\begin{array}}
\newcommand{\ea}{\end{array}}
\newcommand{\bc}{\begin{center}}
\newcommand{\ec}{\end{center}}
\newcommand{\bmat}{\left(\ba}
\newcommand{\emat}{\ea\right)}
\newcommand{\bds}{\begin{description}}
\newcommand{\eds}{\end{description}}
\newcommand{\refs}[1]{(\ref{#1})}
\newcommand{\ler}{\stackrel{\scriptstyle <}{\scriptstyle\sim}}
\newcommand{\ger}{\stackrel{\scriptstyle >}{\scriptstyle\sim}}
\newcommand{\lag}{\langle}
\newcommand{\rag}{\rangle}
\newcommand{\ns}{\normalsize}
\newcommand{\cm}{{\cal M}}
\newcommand{\gr}{m_{3/2}}
\newcommand{\p}{\partial}
\newcommand{\bsg}{$b\rightarrow s + \g$}
\newcommand{\Bsg}{$B\rightarrow X_s + \g$}
\newcommand{\atal}{{\it et al.}}
\newcommand{\cq}{{\cal Q}}
\newcommand{\cqt}{{\widetilde {\cal Q}}}
\newcommand{\wtlc}{{\widetilde C}}
\newcommand{\tcr}{\textcolor{red}}
\def\321{$SU(3)\times SU(2)\times U(1)$}
\def\tl{{\tilde{l}}}
\def\tL{{\tilde{L}}}
\def\bd{{\overline{d}}}
\def\tL{{\tilde{L}}}
\def\a{\alpha}
\def\b{\beta}
\def\bsg{$ b \rightarrow s + \g$}
\def\g{\gamma}
\def\c{\chi}
\def\d{\delta}
\def\D{\Delta}
\def\db{{\overline{\delta}}}
\def\Db{{\overline{\Delta}}}
\def\e{\epsilon}
\def\f{\frac}
\def\tn{-\frac{2}{9}}
\def\tt{\frac{2}{3}}
\def\l{\lambda}
\def\n{\nu}
\def\m{\mu}
\def\nt{{\tilde{\nu}}}
\def\p{\phi}
\def\P{\Phi}
\def\k{\kappa}
\def\x{\xi}
\def\r{\rho}
\def\s{\sigma}
\def\t{\tau}
\def\th{\theta}
\def\ne{\nu_e}
\def\nm{\nu_{\mu}}
\def\snui{\tilde{\nu_i}}
\def\la{{\makebox{\tiny{\bf loop}}}}
\def\ti{\tilde}
\def\ssc{\scriptscriptstyle}
\def\wtl{\widetilde}
\def\mp{\marginpar}
\def\und{\underline}
\def\vckm{V_{\!\mbox{\tiny CKM}}}
\renewcommand{\Huge}{\Large}
\renewcommand{\LARGE}{\Large}
\renewcommand{\Large}{\large}


\preprint{{\vbox{\hbox{NCU-HEP-k026} \hbox{Nov 2006} \hbox{ed. Apr 2007} }}}

\vspace*{1in}
\title{Quark Loop Contributions to Neutron, Deuteron, and Mercury
EDMs from Supersymmetry without R parity}

\vspace*{.5in}

\author{Chan-Chi Chiou}
\author{Otto C. W. Kong}
\email{otto@phy.ncu.edu.tw}
\affiliation{Department of Physics, National Central University,
Chung-Li, TAIWAN 32054.}
\author{Rishikesh D. Vaidya}
\email{rishidilip@gmail.com}
\affiliation{Department of Theoretical Physics\\
Tata Institute of Fundamental Research, Mumbai 400005, India}

\vspace*{1in}

\begin{abstract}
  We present a detailed analysis together with numerical calculations
  on one-loop contributions to the neutron, deuteron and mercury
  electric dipole moment from supersymmetry without R parity, focusing
  on the quark-scalar loop contributions. Being proportional to top
  Yukawa and top mass, such contributions are often large, and since
  these are proportional to hitherto unconstrained combinations of
  bilinear and trilinear RPV parameters, they are all the more
  interesting. Complete formulas are given for the various
  contributions through the quark dipole operators including the
  contribution from color dipole operator. The contribution from color
  dipole operator is found to be similar order in magnitude when
  compared to the electric dipole operator and should be included in
  any consistent analysis.  Analytical expressions illustrating the
  explicit role of the R-parity violating parameters are given
  following perturbative diagonalization of mass-squared matrices for
  the scalars. Dominant contributions come from the combinations
  $B_i^{\ast} \l^{\prime}_{ij1}$ for which we obtain robust bounds. It
  turns out that neutron and deuteron EDMs receive much stronger
  contributions than mercury EDM and any null result at the future
  deuteron EDM experiment or Los Alamos neutron EDM experiment can
  lead to extra-ordinary constraints on RPV parameter space.  Even if
  R-parity violating couplings are real, CKM phase does induce RPV
  contribution and for some cases such a contribution is as strong as
  contribution from phases in the R-parity violating couplings.
  Hence, we have bounds directly on $|B_i^{\ast} \l^{\prime}_{ij1}|$
  even if the RPV parameters are all real. Interestingly, even if
  slepton mass and/or $\mu_{\ssc 0}$ is as high as 1 TeV, it still
  leads to neutron EDM that is an order of magnitude larger than the
  sensitivity at Los Alamos experiment. Since the results are not
  much sensitive to $\tan \beta$, our constraints will survive even if
  other observables tighten the constraints on $\tan \beta$.
\end{abstract}
\pacs{..}
\keywords{neutron EDM, R-parity Violation, Supersymmetry}
\maketitle

\newpage
\section{Introduction}
The problems of neutrino mass, baryogenesis, dark matter, dark energy
and gauge hierarchy, provide unambiguous hints toward physics beyond
standard model (SM). Whereas a direct discovery of new physics
particles at colliders is indispensable, a search for alternative
observables could not only provide a means to discovery of new physics
but also prove complimentary by hinting at favorable regions in
parameter space. Discrete symmetries and their violations have been
crucial to establishing and validating the SM.  Forty years after the
discovery of CP-violation \cite{cpv}, its experimentally observed
effects in the K and B-meson systems
\cite{cpv,KBcpv} are generally compatible with the standard model (SM)
predictions with the Kobayashi-Maskawa (KM) phase as its sole
source. The search for more CP violating observables is keenly pursued
at ongoing and upcoming B physics experiments, and it is largely
confined to flavor changing sector. Within the flavor diagonal sector,
P and T violating electric dipole moments (EDMs) of fermions
\cite{Landaucpv}, heavy atoms and molecules are interesting CP
violating observables that provide essentially background free and
sensitive probes of physics beyond SM
\cite{pospelov-review,ginges-review}. Though the search for
non-vanishing EDM has so far yielded null results, the present
experimental scenario with regard to EDM measurements is very
encouraging, with most of them well within the range of interesting
predictions from physics beyond SM. For the convenience of reader,
below we briefly describe the current status of EDM experiments. This
will also serve to motivate our case for specific new physics
contributions that we discuss here.

Since the early work of Purcell and Ramsey \cite{purcell}, EDM
experiments have dramatically improved in precision. The standard
method for measuring a permanent EDM of a particle is by placing it
in an external electric field $E$ and looking for a shift in energy
that is linear in  $E$. This explains the focus on the electrically
neutral candidates for EDM measurements. Naively, the Schiff
screening theorem prevents any measurement of atomic EDM. Indeed,
the assumptions of Schiff theorem, namely the point-sized nucleus
and non-relativistic limit, are significantly violated in heavy
atoms and hence facilitate EDM measurement. Ironically, the
shielding which is complete in non-relativistic limit, actually
produces an enhancement in the realistic relativistic limit
\cite{salpeter-58}. Sanders \cite{sanders-65} pointed out that due
to relativistic magnetic effects, the atomic EDM induced in heavy
atoms can be strongly enhanced compared to electron EDM inducing it,
leading to the best limit on electron EDM. The enhancement factor
can be over two orders of magnitude. For the case of paramagnetic
atoms, the best bound so far is for the thallium
($~^{205}\mathrm{Tl}$), $|d_{\mathrm{Tl}}| < 9.0 \times 10^{-25}$
(90\%C.L.) leading to the tightest limit on electron EDM $|d_e| \leq
1.6 \times 10^{-27}$ \cite{Tl-bound}. Note that the numbers are in
the standard  $e$ cm unit, which is assumed throughout the paper.
For the case of diamagnetic atoms, mercury ($~^{199}\mathrm{Hg}$)
EDM is best constrained by the Washington group, the bound being,
$|d_{\mathrm{Hg}}| < 2 \times 10^{-28}$ (95 \% C.L.)
\cite{Hg-bound}. An upgraded experiment is expected to improve the
accuracy by a factor of four \cite{Hg-upgrade}. The violation of
Schiff theorem comes about due to the finite size effect of the
nucleus. For the nucleon EDM, the best bound so far is on the
neutron EDM from the Grenoble experiment, $|d_n| < 6.3 \times
10^{-26}$ (90\% C.L.) \cite{expNEDM}. The most recent result from
the same experiment is $|d_n| < 3 \times 10^{-26}$
\cite{nEDM-recent}. It is expected to reach a goal of $|d_n| < 1.5
\times 10^{-26}$ \cite{nEDM-goal}. The Los Alamos neutron EDM
experiment will provide two orders of magnitude improvement, probing
$d_n$ down to order $10^{-28}$ \cite{nEDM-SNS}. The standard method
of measuring the energy shift linear in electric field fails for the
EDM of charged particle due to acceleration of charged particle.
However, in recent years a new dedicated method of searching for EDM
of charged particles in storage rings has been developed
\cite{farley-PRL,semer-00}. A muon EDM experiment is proposed that
is expected to reach a sensitivity of $10^{-24}$ which is an
improvement of a factor of $10^{5}$ to $10^{6}$ over the last CERN
muon g-2 experiment \cite{jpark-letter}. A deuteron EDM experiment,
again using storage ring is proposed, that would reach a sensitivity
of $10^{-27}$  that is 10 to 100 times better than current EDM
limits in terms of sensitivity to quark EDM and QCD $\theta$
parameter \cite{dt-edm-proposal}.

With so many exciting ongoing and upcoming experiments, EDM searches
certainly provide complementary alternative to probe physics beyond
the SM.  In this work we will focus on the neutron, deuteron and
mercury EDM. As an example of physics beyond SM, we shall focus on
supersymmetry (SUSY) without R-parity or the generic supersymmetric
standard model \cite{GSSM}. When the large number of baryon or
lepton number violating terms are removed by imposing an {\it ad
hoc} discrete symmetry called R-parity, one obtains the MSSM
Lagrangian. GSSM is a complete theory of SUSY without R-parity,
where all kinds of R-parity violating (RPV) terms are admitted
without bias. It is generally better motivated than {\it ad hoc}
versions of RPV theories. The MSSM itself, without extension such as
adding SM singlet superfields and admitting violation of lepton
number, cannot accommodate neutrino mass mixings and hence
oscillations \cite{neutrino}. Given SUSY, the GSSM is actually
conceptually the simplest framework to accommodate the latter. The
large number of {\it a priori} arbitrary RPV terms do make
phenomenology complicated. However, the origin of the (pattern of)
values for the couplings may be considered to be on the same footing
as that of the SM Yukawa couplings. For example, it has been shown
in \cite{u1} that one can understand the origin, pattern and
magnitude of all the RPV terms as a result of a spontaneously broken
anomalous Abelian family symmetry model.

As we will see in the next section, EDM of mercury, deuteron and
neutron are expressible in terms of quark EDM and color EDM (CEDM).
Neutron EDM is an old favorite and within SM, the CKM phase
contribution starts at three loop level \cite{shabalin}. Generic
(R-parity conserving) SUSY contributions starts at one loop level
and hence can be large
\cite{arnowitt-D90,oshimo-D92,nath-D98,oshimo-D97,dp-D01,olive-B98,
  hisano-D04,pospelov-ritz-D01,farzan-J05,luca-05,farzan-06}.
Most of the phenomenology studies of the case admitting R-parity
violation have largely been confined to the trilinear superpotential
parameters. Within the latter framework, it has been shown that contributions to fermion 
EDMs start at two loop level  \cite{2loop}. Striking one-loop contributions are, however, 
identified and discussed based on the GSSM framework  \cite{nedm,chun}. Under that 
generic setting, all RPV couplings are considered without bias. Note that studies of RPV 
physics under such a generic setting is uncommon. Not only that {\em a priori} 
approximations  were usually taken on the form of R-parity violation especially 
on the set of bilinear parameters, such approximations were often times not clearly 
stated, if appreciated well enough. They may be confused with the flavor basis choice 
issues (see Ref.\cite{GSSM} for detailed discussions on the aspects). EDM study from 
Godbole {\it el.al} in Ref.\cite{2loop}, however, did state explicitly the bilinear 
couplings were neglected in their study. The one loop contributions are exactly
resulted from combinations of a bilinear and a trilinear RPV couplings.
Similarly, flavor off-diagonal dipole moment contributing to the case of the
$b\rightarrow s \g$ decay \cite{bsg} and that of the $\m \rightarrow
e \g$ \cite{mueg} decay at one-loop level have also been presented.
That is the approach taken here. In Ref.\cite{nedm}, the one loop
EDM contributions from the gluino loop, chargino-like loop,
neutralino-like loop are studied numerically in some details. It
shows that the RPV parameter combination
$\mu_i^*\lambda^{\!\prime}_{i\scriptscriptstyle 1\!1}$ dominates.
with small sensitivity to the value of $\tan\!\beta$. The 
experimental bound on neutron EDM is used to constrain the model
parameter space, especially the RPV part.  However, the alternative
one-loop contribution containing a quark and a scalar in the loop is
also important. For the case of down quark EDM, it contains a
top-quark loop and hence proportional to top mass and top Yukawa,
thus can giving rise to very large contribution. We give complete
1-loop formulas for these contributions to the EDMs of the up- and
down-sector quarks, with full incorporation of family mixings (in
Ref.\cite{nedm}, family mixing was neglected for simplicity). We
present numerical analysis of quark-scalar loop contributions from
all possible combinations of RPV parameters. Besides the familiar
$\mu_i^*\lambda^{\!\prime}_{i\scriptscriptstyle
  1\!1}$, there are a list of combinations of the type
$B_i^*\lambda^{\!\prime}_{ijk}$ which are particularly interesting
and will be the focus of this paper. In comparison to the earlier
works \cite{nedm,chun}, our investigation  provides more extensive
results, both analytically as well as numerically with elaboration
on some physics issues as well as includes contributions to deuteron
and mercury EDM. We obtain bounds on combinations of RPV couplings
which are otherwise unavailable. In fact, as we will see, any null
results of future EDM experiments lead to stringent constraints on
RPV parameters that have not been constrained so far.

Note that we decouple any explicit discussion of leptonic EDM, like
for the electron and muon as well as $d_{\mathrm{Tl}}$ in which the
electron EDM has a dominant role, from the study here. Unlike the
case of the R-parity conserving contributions, the RPV contributions
to EDMs of the quark and lepton sectors have quite independent
origin. For the lepton sector, it involves the $\lambda$-couplings,
rather than the $\lambda^{\!\prime}$-couplings. Hence, a whole set
of different combinations of RPV parameters are to be constrained by
the leptonic EDMs numbers --- not to be addressed in this paper.

The structure of the paper is as follows. In the next section we
describe our notation and framework, the so called single-VEV
parametrization (SVP). We will also describe the formula for
neutron, deuteron and mercury EDM in terms of quark EDM and
chromoelectric dipole moment (CEDM). In section III, we will
describe the quark loop contribution to quark EDM and CEDM coming
from a combination of a bilinear and trilinear RPV couplings. In
section IV we will discuss the results and in section V we will
conclude.

\section{Formulation and Notation}
We summarize the model here while setting the notation. Details of the
formulation adopted is elaborated in Ref.\cite{GSSM}.  The most
general renormalizable superpotential for the supersymmetric SM
(without R-parity) can be written as
\small\begin{eqnarray}
W \!\! &=& \!\varepsilon_{ab}\Big[ \mu_{\alpha}  \hat{H}_u^a \hat{L}_{\alpha}^b
+ h_{ik}^u \hat{Q}_i^a   \hat{H}_{u}^b \hat{U}_k^{\scriptscriptstyle C}
+ \lambda_{\alpha jk}^{\!\prime}  \hat{L}_{\alpha}^a \hat{Q}_j^b
\hat{D}_k^{\scriptscriptstyle C}
\nonumber \\
&+&
\frac{1}{2}\, \lambda_{\alpha \beta k}  \hat{L}_{\alpha}^a
 \hat{L}_{\beta}^b \hat{E}_k^{\scriptscriptstyle C} \Big] +
\frac{1}{2}\, \lambda_{ijk}^{\!\prime\prime}
\hat{U}_i^{\scriptscriptstyle C} \hat{D}_j^{\scriptscriptstyle C}
\hat{D}_k^{\scriptscriptstyle C}  \; ,
\end{eqnarray}\normalsize
where  $(a,b)$ are $SU(2)$ indices, $(i,j,k)$ are the usual family (flavor)
indices, and $(\alpha, \beta)$ are extended flavor index going from $0$ to $3$.
In the limit where $\lambda_{ijk}, \lambda^{\!\prime}_{ijk},
\lambda^{\!\prime\prime}_{ijk}$ and $\mu_{i}$  all vanish,
one recovers the expression for the R-parity preserving case, with
$\hat{L}_{0}$ identified as $\hat{H}_d$. Without R-parity imposed, the
latter is not {\it a priori} distinguishable from the $\hat{L}_{i}$'s.
Note that $\lambda$ is antisymmetric in the first two indices, as
required by the $SU(2)$ product rules, as shown explicitly here with
$\varepsilon_{\scriptscriptstyle 12} =-\varepsilon_{\scriptscriptstyle
21}=1$.  Similarly, $\lambda^{\!\prime\prime}$ is antisymmetric in the
last two indices, from $SU(3)_{\scriptscriptstyle C}$.

The large number of new parameters involved, however, makes the theory
difficult to analyze.  An optimal parametrization, called the
single-VEV parametrization (SVP) has been advocated\cite{ru2} to make
the the task manageable. Here, the choice of an optimal
parametrization mainly concerns the 4 $\hat{L}_\alpha$ flavors. Under
the SVP, flavor bases are chosen such that : 1) among the
$\hat{L}_\alpha$'s, only $\hat{L}_0$, bears a VEV, {\it i.e.} {\small
$\langle \hat{L}_i \rangle \equiv 0$}; 2) {\small $h^{e}_{jk} (\equiv
\lambda_{0jk}) =\frac{\sqrt{2}}{v_{\scriptscriptstyle 0}} \,{\rm diag}
\{m_{\scriptscriptstyle 1},
m_{\scriptscriptstyle 2},m_{\scriptscriptstyle 3}\}$}; 3) {\small
$h^{d}_{jk} (\equiv \lambda^{\!\prime}_{0jk} =-\lambda_{j0k}) =
\frac{\sqrt{2}}{v_{\scriptscriptstyle 0}}{\rm diag}\{m_d,m_s,m_b\}$};
4) {\small $h^{u}_{ik}=\frac{\sqrt{2}}{v_{\scriptscriptstyle u}}
V_{\!\mbox{\tiny CKM}}^{\!\scriptscriptstyle T}\; {\rm
diag}\{m_u,m_c,m_t\}$}, where ${v_{\scriptscriptstyle 0}} \equiv
\sqrt{2}\,\langle \hat{L}_0 \rangle$ and ${v_{\scriptscriptstyle u} }
\equiv \sqrt{2}\,
\langle \hat{H}_{u} \rangle$.
The big advantage of here is that the (tree-level) mass matrices for
{\it all} the fermions {\it do not} involve any of the trilinear RPV
couplings, though the approach makes {\it no assumption} on any RPV
coupling including even those from soft SUSY breaking; and all the
parameters used are uniquely defined, with the exception of some
possibly removable phases.

The soft SUSY breaking part of the Lagrangian in GSSM can be written as follows :
\begin{eqnarray}
V_{\rm soft} &=& \epsilon_{\!\scriptscriptstyle ab}
  B_{\za} \,  H_{u}^a \tilde{L}_\za^b +
\epsilon_{\!\scriptscriptstyle ab}
\left[ \, A^{\!\scriptscriptstyle U}_{ij} \,
\tilde{Q}^a_i H_{u}^b \tilde{U}^{\scriptscriptstyle C}_j
+ A^{\!\scriptscriptstyle D}_{ij}
H_{d}^a \tilde{Q}^b_i \tilde{D}^{\scriptscriptstyle C}_j
+ A^{\!\scriptscriptstyle E}_{ij}
H_{d}^a \tilde{L}^b_i \tilde{E}^{\scriptscriptstyle C}_j   \,
\right] + {\rm h.c.}\nonumber \\
&+&
\epsilon_{\!\scriptscriptstyle ab}
\left[ \,  A^{\!\scriptscriptstyle \lambda^\prime}_{ijk}
\tilde{L}_i^a \tilde{Q}^b_j \tilde{D}^{\scriptscriptstyle C}_k
+ \frac{1}{2}\, A^{\!\scriptscriptstyle \lambda}_{ijk}
\tilde{L}_i^a \tilde{L}^b_j \tilde{E}^{\scriptscriptstyle C}_k
\right]
+ \frac{1}{2}\, A^{\!\scriptscriptstyle \lambda^{\prime\prime}}_{ijk}
\tilde{U}^{\scriptscriptstyle C}_i  \tilde{D}^{\scriptscriptstyle C}_j
\tilde{D}^{\scriptscriptstyle C}_k  + {\rm h.c.}
\nonumber \\
&+&
 \tilde{Q}^\dagger \tilde{m}_{\!\scriptscriptstyle {Q}}^2 \,\tilde{Q}
+\tilde{U}^{\dagger}
\tilde{m}_{\!\scriptscriptstyle {U}}^2 \, \tilde{U}
+\tilde{D}^{\dagger} \tilde{m}_{\!\scriptscriptstyle {D}}^2
\, \tilde{D}
+ \tilde{L}^\dagger \tilde{m}_{\!\scriptscriptstyle {L}}^2  \tilde{L}
  +\tilde{E}^{\dagger} \tilde{m}_{\!\scriptscriptstyle {E}}^2
\, \tilde{E}
+ \tilde{m}_{\!\scriptscriptstyle H_{\!\scriptscriptstyle u}}^2 \,
|H_{u}|^2
\nonumber \\
&& + \frac{M_{\!\scriptscriptstyle 1}}{2} \tilde{B}\tilde{B}
   + \frac{M_{\!\scriptscriptstyle 2}}{2} \tilde{W}\tilde{W}
   + \frac{M_{\!\scriptscriptstyle 3}}{2} \tilde{g}\tilde{g}
+ {\rm h.c.}\; ,
\label{soft}
\end{eqnarray}
where we have separated the R-parity conserving ones from the
RPV ones ($H_{d} \equiv \hat{L}_0$) for the $A$-terms. Note that
$\tilde{L}^\dagger \tilde{m}_{\!\scriptscriptstyle \tilde{L}}^2  \tilde{L}$,
unlike the other soft mass terms, is given by a
$4\times 4$ matrix. Explicitly,
$\tilde{m}_{\!\scriptscriptstyle {L}_{00}}^2$ is
$\tilde{m}_{\!\scriptscriptstyle H_{\!\scriptscriptstyle d}}^2$
of the MSSM case while
$\tilde{m}_{\!\scriptscriptstyle {L}_{0k}}^2$'s give RPV mass mixings.

Details of the tree-level mass matrices for all fermions and scalars
are summarized in Ref.\cite{GSSM}. For the analytical appreciation of
many of the results, approximate expressions of all the RPV mass
mixings are very useful. The expressions are available from
perturbative diagonalization of the mass matrices\cite{GSSM}.

\section{The Quark Loop Contribution to neutron, deuteron and Mercury
  EDM}
As we will see toward the end of this section, neutron, deuteron,
and mercury EDMs are ultimately expressible in terms of quark EDMs
and CEDMs. Quark EDMs and CEDMs are typically defined from the
following effective Lagrangian: 
\be {\cal L}_{\mathrm{eff.}} = -\f{i}{2}d^{\!\ssc E}_f \bar{f}
\sigma_{\m\n}\g_5 f F^{\m\n} -\f{i}{2} d^{\!\ssc C}_f \bar{f}
\sigma_{\m\n}\g_5 T^af G^{\m\n a} \;. \ee 
Here, $d^{\!\ssc E}_f$ and $d^{\!\ssc C}_f$ are EDM and CEDM,
respectively, of a quark flavor $f$. We perform calculations of the
one-loop EDM diagrams using mass eigenstates with their effective
couplings. The approach frees our numerical results from the
mass-insertion approximation more commonly adopted in the type of
calculations, while analytical discussions based of the perturbative
diagonalization formulae help to trace the major role of the RPV
couplings, especially those of the bilinear type.  The interesting class of one-loop 
contributions are obtained from diagrams involving a bilinear-trilinear parameter 
combinations, which were seldom studied by most other authors on RPV physics.
 The bilinear parameters come into play through mass
mixings induced among the fermions and scalars (slepton and Higgs
states), while the trilinear parameter enters a effective coupling
vertex.  The basic features are the same as those reported in the
studies of the various related processes\cite{nedm,bsg}. Among the
latter, our recently available reports on \bsg\ \cite{bsg} are
particularly noteworthy, for comparison. The $d$ quark dipole plays
a more important role over that of the $u$ quark, when RPV
contributions are involved. The \bsg\ diagram is, of course, nothing
other than a flavor off-diagonal version of a down-sector quark
dipole moment diagram.

\begin{figure}
\includegraphics[scale=1.0]{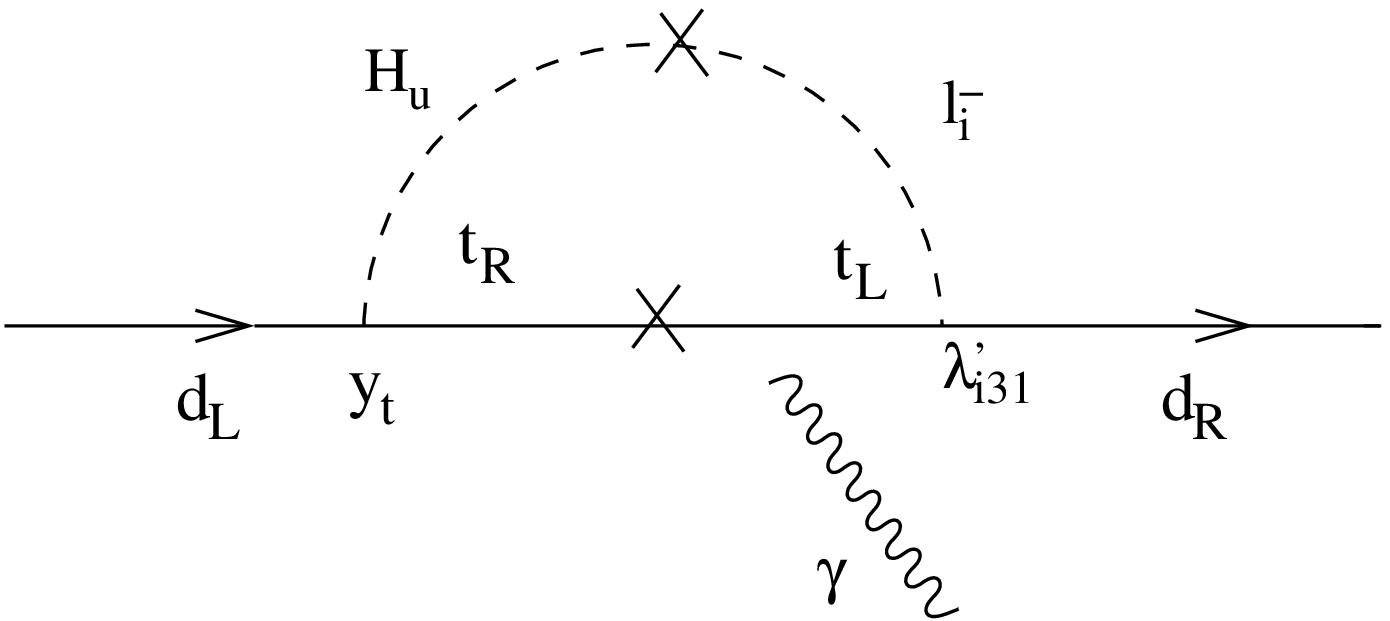}
\caption{$d$ quark EDM due to $B^{\ast}_i
\l^{\prime}_{i31}$ combination. Due to top Yukawa and top-mass
dependence this is the most dominant contribution.}
\label{dedm}
\end{figure}

Unlike the SM Yukawa couplings, the RPV Yukawa couplings 
are mostly not so strongly constrained in magnitudes \cite{bounds}, and
are sources of flavor mixings. A trilinear $\lambda_{\za
jk}^{\!\prime}$ coupling couples a quark to another one, of the same
or different charge, and a generic scalar, neutral or charged
accordingly.  The coupling together with a SM Yukawa coupling at
another vertex contributes to the EDMs through some scalar mass
eigenstates with RPV mass mixings involved, as illustrated in Fig.
\ref{dedm}. As the SM Yukawas are flavor diagonal, the charged
scalar loops contribute by invoking CKM mixings\footnote{ For the
corresponding flavor off-diagonal transition moment, like \bsg\ ,
scalar loop contributions do exist\cite{bsg}.}.  Note that under the
SVP adopted, the $\lambda_{\za jk}^{\!\prime}$ parameters have quark
flavor indices actually defined in the $d$-sector quark mass
eigenstate basis. Analytically, we have the formula for the electric
dipole form factor for quark flavor $f$, 
\bea \small \label{eq:edm-formula} 
\left( 
{\frac{d^{\!\ssc E}_{\scriptscriptstyle f}}{e}}\right)_{\!\!\phi
^{\mbox{-}}}  =   -{ \frac{\alpha _{\!\mbox{\tiny em}}}{4\pi \,\sin
\!^{2}\theta _{\! \scriptscriptstyle W}}}\;  \sum_{m}^{\prime
}\sum_{n=1}^{3}\,\mbox{Im}( \wtl{\cal C}_{\ssc nmi}^{\ssc
\!L}\wtl{\cal C}^{\ssc \!R \ast}_{\ssc nmi})
\frac{{M}_{\!\scriptscriptstyle f_{n}^{\prime }}}{M_{\!\scriptscriptstyle
\tilde{\ell}_{m}}^{2}}\;
\left[ ({\cal Q}_{\!f}-{\cal Q}_{\!f^{\prime
}})\;F_{4}\!\left( {\frac{{M}_{\!\scriptscriptstyle f_{n}^{\prime }}^{2}}{
M_{\!\scriptscriptstyle\tilde{\ell}_{m}}^{2}}}\right) -{\cal Q}_{\!f^{\prime
}}\;F_{3}\!\left( {\frac{{M}_{\!\scriptscriptstyle f_{n}^{\prime }}^{2}}{
M_{\!\scriptscriptstyle\tilde{\ell}_{m}}^{2}}}\right) \right] \;,
\end{eqnarray} 
and for the chromo-electric dipole form factor, 
\be   \label{eq:cedm-formula} \left( 
d^{\!\ssc C}_{\scriptscriptstyle f}\right) _{\!\!\phi ^{\mbox{-}}}={
\frac{g_s \alpha _{\!\mbox{\tiny em}}}{4\pi \,\sin \!^{2}\theta _{\!
\scriptscriptstyle W}}}\;\sum_{m}^{\prime
}\sum_{n=1}^{3}\,\mbox{Im}( \wtl{\cal C}_{\ssc nmi}^{\ssc
\!L}\wtl{\cal C}^{\ssc \!R \ast}_{\ssc nmi})
\frac{{M}_{\!\scriptscriptstyle f_{n}^{\prime }}}{M_{\!\scriptscriptstyle
\tilde{\ell}_{m}}^{2}}\;
{\cal Q}_{\!f^{\prime
}}\;F_{3}\!\left( {\frac{{M}_{\!\scriptscriptstyle f_{n}^{\prime }}^{2}}{
M_{\!\scriptscriptstyle\tilde{\ell}_{m}}^{2}}}\right)  \;.
\ee
Here, $F_3 (x)$ and $F_4 (x)$ are Inami-Lim loop functions, given as:
\bea
F_3(x) &=&
\frac{1}{2(1-x)^3} \, (-3+4 \, x-x^2-2 \,  \ln x) \; ,\\
F_4(x) &=& \frac{1}{2(1-x)^3} \, (1-x^2+2 \, x \,\ln x) \; . 
\eea 
For $f=u$  ($i=1$), the coefficients
$\wtl{\cal C}_{\ssc nmi}^{\! \ssc L,\ssc R}$ are defined by the interaction Lagrangian,
\be
{\cal L}^{u} = {g_{\scriptscriptstyle 2}}
\overline{\Psi}(d_n)\Phi^{\dagger}(\phi_m^{{\mbox{-}}})
\left[
\widetilde{\cal C}^{\!\scriptscriptstyle L}_{\scriptscriptstyle nmi}
{1 - \gamma_{\scriptscriptstyle 5} \over 2}  +
\widetilde{\cal C}^{\!\scriptscriptstyle R}_{\scriptscriptstyle nmi}
{1 + \gamma_{\scriptscriptstyle 5} \over 2}  \right]
{\Psi}({u_{i}})
\ + \mbox{h.c.}\;,
\ee
\begin{eqnarray}
\wtl{\cal C}_{\ssc nmi}^{\ssc L \ast} &=&\frac{y_{d_{n}}}{g_{%
\scriptscriptstyle 2}}\,V_{\!\mbox{\tiny CKM}}^{in}{\cal D}_{2m}^{l^{\ast }}+\frac{\lambda
_{jkn}^{\!\prime \ast }}{g_{\scriptscriptstyle2}}\,V_{\!\mbox{\tiny CKM}}^{ik}{\cal D}%
_{(j+2)m}^{l^{\ast }}\;,  \nonumber \\
\wtl{\cal C }_{\ssc nmi}^{\ssc R \ast} &=&\frac{y_{u_{i}}}{g_{%
\scriptscriptstyle 2}}V_{\!\mbox{\tiny CKM}}^{in}\,{\cal D}_{1m}^{l^{\ast }}\;,
\end{eqnarray}
and, for $f=d$  ($i=1$), the coefficients
$\wtl{\cal C}_{\ssc nmi}^{\! \ssc L,\ssc R}$ are defined by the interaction Lagrangian,
\be
{\cal L}^{d} = {g_{\scriptscriptstyle 2}}
\overline{\Psi}(u_n)\Phi (\phi_m^{{\mbox{-}}})
\left[
\widetilde{\cal C}^{\!\scriptscriptstyle L}_{\scriptscriptstyle nmi}
{1 - \gamma_{\scriptscriptstyle 5} \over 2}  +
\widetilde{\cal C}^{\!\scriptscriptstyle R}_{\scriptscriptstyle nmi}
{1 + \gamma_{\scriptscriptstyle 5} \over 2}  \right]
{\Psi}({d_{i}})
\ + \mbox{h.c.}\;,
\ee
\begin{eqnarray}
\wtl{\cal C}_{\ssc nmi}^{\ssc L \ast}
&=&\frac{y_{u_{n}}}{g_{%
\scriptscriptstyle 2}}V_{\!\mbox{\tiny CKM}}^{ni^{\ast }}\,{\cal D}_{1m}^{l}\;,
\nonumber \\
\wtl{\cal C }_{\ssc nmi}^{\ssc R \ast}
&=&\frac{y_{d_{i}}}{g_{%
\scriptscriptstyle 2}}\,V_{\!\mbox{\tiny CKM}}^{ni^{\ast }} {\cal D}^l_{2m}+
\frac{\lambda _{jki}^{\!\prime
}}{g_{\scriptscriptstyle 2}}\,V_{\!\mbox{\tiny CKM}}^{nk^{\ast }}{\cal D}
_{(j+2)m}^{l}\;,
\end{eqnarray}
with the $\sum_{m}^{\prime }$ denoting a sum over (seven) nonzero mass
eigenstates of the charged scalar; i.e., the unphysical Goldstone mode
is dropped from the sum, $D^{l}$ being the diagonalization matrix,
i.e., $D^{l\dag }M_{\!\scriptscriptstyle
E}^{2}\,D^{l}=\mbox{diag}\{\,M_{\!\scriptscriptstyle\tilde{\ell}
_{m}}^{2},m=1\,\mbox{--}\,8\,\}$ and $M_{\!\scriptscriptstyle
E}^{2}$ being $8\times 8 $ charged-slepton and Higgs mass-squared matrix .
Here, $i=1$ is the flavor index for the external quark while $n$ is
for the quark running inside the loop. Note that the unphysical
Goldstone mode is dropped from the scalar sum because it is rather a
part of the gauge loop contribution, which obviously is real and does
not affect the EDM calculation.

In general, there also exist the contributions from neutral scalar
loop (involving the mixing of neutral Higgs with sneutrino in the
loop). These contributions lack the top Yukawa and top mass
effects that enhance the contributions of charged-slepton loop, and
hence less important. The formula for the electric dipole form factor
is given as:
\be
\label{eq:neutral-scalar}
\left( {\frac{d^E_{\scriptscriptstyle
f}}{e}}\right)_{\!\!\phi^{\ssc 0}}=
-
\frac{\alpha _{\!\mbox{\tiny em}} {\cal Q}_f}{4\pi \,\sin \!^{2}\theta _{\!
\scriptscriptstyle W}}\;\sum_{m}^{\prime }\sum_{n=1}^{3}\,\mbox{Im}(
\wtl{\cal N}_{\ssc nmi}^{\ssc \!L}\wtl{\cal N}^{\ssc \!R \ast}_{\ssc nmi})
\frac{{M}_{\!\scriptscriptstyle f_{n}}}{M_{\!\scriptscriptstyle
\tilde{s}_{m}}^{2}}\;
\;F_{3}\!\left( {\frac{{M}_{\!\scriptscriptstyle f_{n}}^{2}}{
M_{\!\scriptscriptstyle\tilde{\ell}_{m}}^{2}}}\right) \;,
\ee
and for the chromo-electric dipole form-factor is given as,
\be
\left( d^C_{\scriptscriptstyle f} \right)_{\!\!\phi^{\ssc 0}}=
-\frac{g_s\alpha _{\!\mbox{\tiny em}} {\cal Q}_f}{4\pi \,\sin \!^{2}\theta _{\!
\scriptscriptstyle W}}\;\sum_{m}^{\prime }\sum_{n=1}^{3}\,\mbox{Im}(
\wtl{\cal N}_{\ssc nmi}^{\ssc \!L}\wtl{\cal N}^{\ssc \!R \ast}_{\ssc nmi})
\frac{{M}_{\!\scriptscriptstyle f_{n}}}{M_{\!\scriptscriptstyle
\tilde{s}_{m}}^{2}}\;
\;F_{3}\!\left( {\frac{{M}_{\!\scriptscriptstyle f_{n}}^{2}}{
M_{\!\scriptscriptstyle\tilde{\ell}_{m}}^{2}}}\right) \;,
\ee
where, for $f=u$  ($i=1$), the coefficients $\wtl{\cal N}_{\ssc nmi}^{\! \ssc L,\ssc R}$
are defined by the interaction Lagrangian,
\be
{\cal L}^{u} = {g_{\scriptscriptstyle 2}}
\overline{\Psi}(u_n)\Phi^{\dagger}(\phi_m^{{\ssc 0}})
\left[
\widetilde{\cal N}^{\!\scriptscriptstyle L}_{\scriptscriptstyle nmi}
{1 - \gamma_{\scriptscriptstyle 5} \over 2}  +
\widetilde{\cal N}^{\!\scriptscriptstyle R}_{\scriptscriptstyle nmi}
{1 + \gamma_{\scriptscriptstyle 5} \over 2}  \right]
{\Psi}({u_{i}})
\ + \mbox{h.c.}\;,
\ee
\bea
\widetilde{\cal N}^{\!\scriptscriptstyle L \ast}_{\scriptscriptstyle
nmi} & = & -\f{y_{u_i}}{g_2}\d_{in}\f{1}{\sqrt 2} \left[ {\cal
D}^s_{1m} - i {\cal D}^s_{6m} \right] \; ,\nonumber\\
\widetilde{\cal N}^{\!\scriptscriptstyle R \ast}_{\scriptscriptstyle
nmi} & = & -\f{y_{u_i}}{g_2}\d_{in}\f{1}{\sqrt 2} \left[ {\cal
D}^s_{1m} + i {\cal D}^s_{6m} \right]\;;
\eea
and, for $f=d$  ($i=1$), the coefficients
$\wtl{\cal N}_{\ssc nmi}^{\! \ssc L,\ssc R}$ are defined by the
interaction Lagrangian,
\be
{\cal L}^{d} = {g_{\scriptscriptstyle 2}}
\overline{\Psi}(d_n)\Phi (\phi_m^{\ssc 0})
\left[
\widetilde{\cal N}^{\!\scriptscriptstyle L}_{\scriptscriptstyle nmi}
{1 - \gamma_{\scriptscriptstyle 5} \over 2}  +
\widetilde{\cal N}^{\!\scriptscriptstyle R}_{\scriptscriptstyle nmi}
{1 + \gamma_{\scriptscriptstyle 5} \over 2}  \right]
{\Psi}({d_{i}})
\ + \mbox{h.c.}\;,
\ee
\bea
\widetilde{\cal N}^{\!\scriptscriptstyle L \ast}_{\scriptscriptstyle
nmi} & = & -\f{y_{u_i}}{g_2}\d_{in}\f{1}{\sqrt 2} \left[ {\cal
D}^s_{2m} - i {\cal D}^s_{7m} \right]
-\f{\l^{\prime \ast}_{jin}}{g_2}\f{1}{\sqrt 2} \left[
{\cal D}^s_{(j+2)m} - i {\cal D}^s_{(j+7)m} \right] \nonumber\\
\widetilde{\cal N}^{\!\scriptscriptstyle R \ast}_{\scriptscriptstyle
nmi} & = & -\f{y_{d_i}}{g_2}\d_{in}\f{1}{\sqrt 2} \left[ {\cal
D}^s_{2m} + i {\cal D}^s_{7m} \right]
-\f{\l^{\prime}_{jni}}{g_2}\f{1}{\sqrt 2} \left[
{\cal D}^s_{(j+2)m} + i {\cal D}^s_{(j+7)m} \right]
\; .
\eea
with the $\sum_{m}^{\prime }$ denoting a sum over (10) nonzero mass
eigenstates of the neutral scalar; i.e., the unphysical Goldstone mode
is dropped from the sum, $D^{s}$ being the diagonalization matrix for
the $10\times 10$ mass matrix for neutral scalars (real and symmetric,
written in terms of scalar and pseudo-scalar parts). Again $i=1$ is the
flavor index for the external quark while $n$ is for the quark running
inside the loop.

Having obtained expression for quark EDMs and CEDMs, the next task
is to connect them to the hadronic and atomic EDMs. This is a
non-trivial step owing to non-perturbative effects of QCD for which
there is no single unambiguous approach. Below we will briefly
describe the neutron, deuteron, and mercury EDM  formulas that we
will use for numerical calculations. For details we refer the reader
to the excellent review articles \cite{pospelov-review,
ginges-review}.

For the case of neutron EDM, there have been three different
approaches with varying degree of sophistication. The simplest of
these is the non-relativistic $SU(6)$ quark model
\cite{arnowitt-D90,oshimo-D92,nath-D98,oshimo-97,luca-05}. There
have also been attempts based on the chiral Lagrangian method
\cite{hisano-D04} and QCD sum rule approach
\cite{pospelov-ritz-D01}. In the chiral Lagrangian approach neutron
EDM is expressed solely in terms of quark CEDM \cite{hisano-D04}: 
\be 
d_n = (1.6 \, d^{\!\ssc C}_u + 1.3\, d^{\!\ssc C}_d + 0.26 \,
d^{\!\ssc C}_s ) \;. \ee 
In the case of QCD sum rule approach one obtains
\cite{pospelov-ritz-D01}: 
\be 
d_n = (1 \pm 0.5)
\f{|\lag\bar{q}q\rag |}{(225 \mathrm{MeV})^3} \times \left[0.55e(d^{\!\ssc C}_d
  + 0.5d^{\!\ssc C}_u ) + 0.7 (d^{\!\ssc E}_d - 0.25d^{\!\ssc E}_u)\right].  
\ee 
However, the two approaches differ substantially in their dependence
on various quark EDMs and and CEDMs and their relative importance
and hence give very different results. For instance, in the chiral
Lagrangian method quark EDMs are completely neglected and only CEDMs
contribute, while both contribute in the QCD sum rule approach.
Moreover, strange quark CEDM contribution is dominant in chiral
Lagrangian method whereas it is neglected in the QCD sum rule
approach. Owing to the large differences in these method we use the
non-relativistic $SU(6)$ quark model and incorporate the quark CEDM
contributions using naive dimensional analysis. Although, this
approach is less sophisticated than the above two methods, it
provides a reasonable and reliable enough estimate for our present
purpose.

In the $SU(6)$ quark model, one associates a non-relativistic
wavefunction to the neutron consists of three constituent quarks
with two spin states. After evaluating the relevant Clebsch-Gordan
coefficients the neutron EDM is written as 
\be \label{vqm} d_n = {1\over 3} \left (4\, d_d -d_u \right) \;, \ee
where 
\be d_q = \eta^{\!\ssc E}d^{\!\ssc E}_q + \eta^{\!\ssc C}
\frac{e}{4\pi} \,d^{\!\ssc C}_q \;. \ee
Here, $d^{\!\ssc E}$ and $d^{\!\ssc C}$ are the contributions to
neutron EDM from the electric and chromoelectric dipole operators,
respectively. $\eta^{\!\ssc E}( \simeq 0.61)$ and $\eta^{\!\ssc C}
(\simeq 3.4)$ are the respective QCD correction factor from
renormalization group evolution \cite{luca-05}. The authors of
Ref.\cite{nath-D98} showed that within MSSM chromoelectric dipole
form factor contribution to neutron EDM is comparable to that from
electric dipole form factor and hence should be included.

Among diamagnetic atoms, mercury atom provides the best limit on
atomic EDM, which is a result of T-odd nuclear interactions. Schiff
screening theorem is violated by finite size effects of the nucleus
and are characterized by Schiff moment $S$ which generates T-odd
electrostatic potential for atomic electrons. Again the principle
approaches in the literature are the QCD sum rule and chiral
Lagrangian method. The QCD sum rule approach gives
\cite{pospelov-hg} 
\be 
d_{H\! g} = - (d^{\!\ssc C}_d - d^{\!\ssc C}_u - 0.012 \, d^{\!\ssc
C}_s )\times 3.2 \cdot 10^{-2} \; e \;. \ee 
The chiral Lagrangian method gives \cite{hisano-B04} 
\be 
d_{H\! g} = - (d^{\!\ssc C}_d - d^{\!\ssc C}_u - 0.0051 \,
d^{\!\ssc C}_s )\times 8.7\times 10^{-3}  \; e \;. \ee 
Again, one can see the differences in the two approaches. We shall
follow the results of Ref.\cite{pospelov-PLB530} and interpret the
bounds on mercury EDM as constraining the combination $|d^{\!\ssc
C}_u - d^{\!\ssc C}_d| < 2 \times 10^{-26}$. As we will soon see in
our result section, constraints coming from mercury EDM are
redundant as neutron and the planned deuteron EDM experiment provide
much stringent constraints.

For the case of deuteron EDM, because of rather transparent nuclear
dynamics, theoretical uncertainty is much smaller. As mentioned in
the introduction, the proposed deuteron EDM experiment would reach a
sensitivity of $(1-3)\times 10^{-27}$ $e\,$cm that is about 10 to
100 times better than than the current EDM limits in terms of
sensitivity to quark EDMs and QCD $\theta$ parameter
\cite{dt-edm-proposal}. Deuteron EDM receives contributions from the
constituents proton and neutron EDMs as well as CP-odd meson nucleon
couplings: 
\be d_D = d_n + d_p + d^{\ssc \pi NN}_D \;.  \ee 
For theoretical predictions again there are two approaches, namely
QCD sum rule \cite{lebedev-PRD70} and the chiral Lagrangian approach
\cite{Hisano-dt}. Authors of Ref.\cite{lebedev-PRD70} show that in
the $SU(2)$ chiral perturbation theory proton and neutron EDMs
exactly cancel each other and hence compute the EDM using QCD sum
rule approach. Authors of Ref.\cite{Hisano-dt} have shown that this
is not true after introducing strange quark and corresponding meson.
However, the two approaches agree to within error-bars regarding the
dominant contributions due to quark CEDMs and following
Ref.\cite{farzan-J05} we take deuteron EDM to be $d_D (d_q,d^{\!\ssc
C}_q) \simeq -e \, (d^{\!\ssc C}_u - d^{\!\ssc C}_d) \,
5^{+11}_{-3}$ and use the best-fit value for our analysis.

\section{Results and Discussions}
From the discussion in previous section it is clear that analysis of
neutron, deuteron, and mercury EDMs boils down to understanding
quark EDMs and CEDMs. Quark EDMs involve violation of CP but not
R-parity. Thus RPV parameters should come in combinations that
conserve R-parity. From an inspection of formulae it is clear that
the contributions from two $\l^{\prime}$ couplings cannot lead to
EDM as these violate lepton number by two units which has to be
compensated by Majorana like mass insertions for neutrino or
sneutrino propagators 
\footnote{A combination $\l^{\prime
*}_{ijk}\l^{\prime}_{ijk}$ is real and hence does not contribute to
EDM}. 
If one is willing to admit more than two $\l^{\prime}$ to be
non-zero, than in principle one can have contributions to EDM at one
loop level from the Majorana like mass insertions but these would be
highly suppressed. With only two RPV couplings, the only other
possibility at one loop level is to have a combination of trilinear
and a bilinear ($\m_i$, $B_i$ or ${\tilde m}^2_{{\ssc L}_{\ssc
0i}}$) RPV couplings in such a way that lepton number is conserved.
However not all the three bilinears mentioned above are independent
as they are related by tadpole relation in the single VEV
parametrization \cite{GSSM} 
\be B_i \tan\! \beta = {\tilde m}^2_{{\ssc L}_{\ssc 0i}} + \m^*_{\ssc
  0}\m_i \;.  \ee
Using the above tadpole equation we eliminate ${\tilde m}^2_{{\ssc
    L}_{\ssc 0i}}$ in favor of $\m_i$ and $B_i$. Contributions from
the combination $\m^{\ast}_i\l^{\prime}_{ijk}$, through squark
loops, had been extensively studied in \cite{nedm} in detail. Here,
we shall focus on the $B^{\ast}_i \l^{\prime}_{ijk}$ kind of
combination as illustrated in Fig.\ref{dedm}.  Such a combination
can contribute through charged-scalar (charged-slepton, charged
Higgs mixing) loop or neutral scalar (sneutrino, neutral Higgs
mixing) loop. The fermions running inside the loops are quarks,
instead of the gluon or a colorless fermion as in the case of the
squark loops.  Before we discuss the numerical results it would be
worthwhile to discuss the analytical results which can then be
compared with numerical results.

Let us focus on RPV part of $ \mbox{Im}(\wtl{\cal C}_{\ssc
nmi}^{\ssc \!L}\wtl{\cal C}^{\ssc \!R \ast}_{\ssc nmi})$, in
Eq.(\ref{eq:edm-formula})  for the case of $d$ quark EDM. It is
given as 
\be \label{cs-d} \mbox{Im}( \wtl{\cal C}_{\ssc nm1}^{\ssc
\!L}\wtl{\cal C}^{\ssc \!R \ast}_{\ssc nm1})_{\ssc RPV}  =
\mbox{Im}\left[
(y_{u_n}\vckm^{n1}{\cal D}^{l\ast}_{1m})~\times~(\l^{\prime}_{jk1}
\vckm^{nk\ast}{\cal D}^{l}_{(j+2)m})\right] \;. \ee 
For the $u$ quark dipole, we have
\be
\label{cs-u}
\mbox{Im}(
\wtl{\cal C}_{\ssc nm1}^{\ssc \!L}\wtl{\cal C}^{\ssc \!R \ast}_{\ssc
nm1})_{\ssc RPV} = \mbox{Im}\left[
(\l^{\prime}_{jkn}\vckm^{1k\ast} {\cal D}^l_{(j+2)m})
~\times~(y_{u}\vckm^{1n} {\cal D}^{l\ast}_{1m}) \right] \;.  \ee 
Interestingly, the entries of the slepton-Higgs diagonalizing matrix
involved are the same in both terms above.  Bilinear RPV terms are
hidden inside the slepton-Higgs diagonalization matrix elements. To
see the explicit dependence on bilinear RPV terms let us look at the
diagonalizing matrix elements of charged-slepton Higgs mass matrix,
${\cal D}^l_{(j+2)m}{\cal D}^{l\ast}_{1m}$, more closely.  When
summed over the index $m$, it of course gives zero, owing to
unitarity. This is possible only for the case of exact mass
degeneracy among the scalars when loop functions factor out in the
sum over $m$ scalars. In fact, the final result involves a double
summation over the $m$ scalar and $n$ fermion (quark) mass
eigenstates. Either the unitarity constraint over the former sum or
the GIM cancellation over the latter predicts a null result whenever
the mass dependent loop functions $F_{4}\!\left(
{\frac{{M}_{\!\scriptscriptstyle f_{n}^{\prime }}^{2}}{
      M_{\!\scriptscriptstyle\tilde{\ell}_{m}}^{2}}}\right)$ and
$F_{3}\!\left( {\frac{{M}_{\!\scriptscriptstyle f_{n}^{\prime }}^{2}}{
      M_{\!\scriptscriptstyle\tilde{\ell}_{m}}^{2}}}\right)$ can be
factored out of the corresponding summation due to mass degeneracy. In
reality however, these elements are multiplied by non-universal loop
functions giving non-zero EDM. Restricting to first order in
perturbation expansion of the mass-eigenstates, ${\cal
  D}^l_{(j+2)m}{\cal D}^{l\ast}_{1m}$ is non-zero for $m=2$ and
$m=j+2$, both giving similar dependence on RPV but with opposite
sign ($m=1$ is the unphysical Goldstone state that is dropped from
the sum here).  Also ${\cal D}^l_{(j+2),2} \sim {\cal
D}^{l\ast}_{1(j+2)}$. For $m=2$ one obtains \cite{GSSM} 
\be
\label{diag1}
{\cal D}^l_{(j+2)2}\;{\cal D}^{l\ast}_{12} \sim \f{ B^{\ast}_j}{M^2_s}
 \times \mathrm{O}(1)
\ee 
Here, $M^2_s$ denotes the difference in the relevant diagonal entries (generic
mass-squared parameter of the order of soft mass scale) in
charged-slepton Higgs mass matrix.
To first order in perturbation expansion there is no contribution to
EDM from $\m_i$. If one goes to second order in perturbation
expansion, one gets a contribution from the term $m = j+5$ which is
given as \cite{GSSM} 
\be \label{mui} {\cal D}^l_{(j+2)(j+5)}\;{\cal
D}^{l\ast}_{1(j+5)} \sim \f{\m_j^{\ast} m_j}{M_s^2} \times
\left[\f{(A^{\ast}_e - \m_{\ssc 0} \tan \b)m_j}{M^2_s} -
\f{\sqrt{2}M_{\ssc W} \sin \b (\m_k \l^{\ast}_{kjj})}{g_2
M^2_s}\right] \ee There are a few important things to be noted here:
\begin{enumerate}
\item Notice that $\m_i$ enter at second order in perturbation
  expansion. Moreover they are accompanied by corresponding
  charged-lepton mass-squared and hence $\m^{\ast}_i
  \l^{\prime}_{ijk}$ contributions are always suppressed (later below
  we will elaborate on this). Also notice that in principle trilinear
  parameter $\l_{kjj}$ also contribute through $\m_k \l^{\ast}_{kjj}$.
  But they have to be present in addition to the trilinear parameter
  $\l^{\prime}_{ijk}$, thus making it a fourth-order effect in
  perturbation and hence negligible. Thus, we will focus on
  $B^{\ast}_i \l^{\prime}_{ijk}$ effects which are interesting.

\item Even if all RPV parameters are real, CKM phase in conjunction
  with real RPV parameters could still induce EDM. As we will soon
  see, this could be sizable.
\item It is clear that $d$ quark EDM receives much larger contribution
  owing to top-Yukawa and proportionality to top mass. There are nine
  trilinear RPV couplings $\l^{\prime}_{ij1}$ that contribute to $d$
  quark EDM. $\l^{\prime}_{i31}$ has the largest impact owing to the
  least CKM suppression. The $m_t$ enhancement feature is not there in 
  the two loop contributions \cite{2loop}.
\item All the 27 trilinear RPV couplings ($\l^{\prime}_{ijk}$)
  contribute to $u$ quark EDM.  However, the absence of enhancement
  from the top mass in the loop and the uniform proportionality to
  up-Yukawa considerably weakens the type of contribution to $u$ quark
  EDM.
\item There is no RPV neutral scalar loop contribution to $u$ quark
  EDM.  However, there are one-loop contributions to $d$ quark EDM
  from the neutral-Higgs sneutrino mixing due to the combination of
  $B_i \l^{\prime}_{i11}$ with Majorana like mass-insertion in the
  loop\footnote{The Majorana like mass-insertion can be considered a
    result of the non-zero $B_i$. It manifest itself in our exact mass
    eigenstate calculation as a mismatch between the corresponding
    scalar and pseudo-scalar parts complex ``sneutrino" state which
    would otherwise have contributions canceling among themselves.
    With the non-zero $B_i$, the involved diagram is one with two
    $\l^{\prime}_{i11}$ coupling vertices and a internal $d$ quark.
    Hence, the type of contribution is possible only with the single
    $\l^{\prime}$ coupling.  }. From the EDM formula, it is clear that
  this would be about the same magnitude as the $u$ quark EDM due to
  charged-Higgs charged slepton mixing and hence highly suppressed.
\end{enumerate}

From the above analytical discussion, we illustrated clearly how
various combinations of trilinear and bilinear RPV parameters
contribute to neutron EDM.  We now focus on the numerical results.
In order to focus on individual contributions we keep a pair of RPV
coupling (a bilinear and a trilinear) to be non-zero at a time to
study its impact. We have chosen all sleptons and down-type Higgs to
be 100 GeV (up-type Higgs mass and $B_{0}$ being determined from
electroweak symmetry breaking conditions), $\m_{\ssc 0}$ parameter
to be -300 GeV, $A$ parameter at the value of 300 GeV, and $\tan\!\b
= 3$ (we will show the impact of some parameter variations below).
Since the $\l^{\prime}$ couplings are on the same footing as the
standard Yukawa couplings, they are in general complex.  We admit a
phase of $\pi/4$ for non-zero $\lambda^{\prime}$ couplings while
still keeping the CKM phase. The phases of the other R-parity
conserving parameters are switched off. Since it is the relative
phase of the $B^{\ast}_i \l^{\prime}_{ijk}$ product that is
important, we put the phase for $B_i$ to be zero without loss of
generality.  We do not assume any hierarchy in the sleptonic
spectrum. Hence, it is immaterial which of the bilinear parameter
$B_i$ is chosen to be non-zero. We choose $B_3$ to be non-zero but
all the results hold good for $B_1$ and $B_2$ as well.

\begin{figure}
\includegraphics{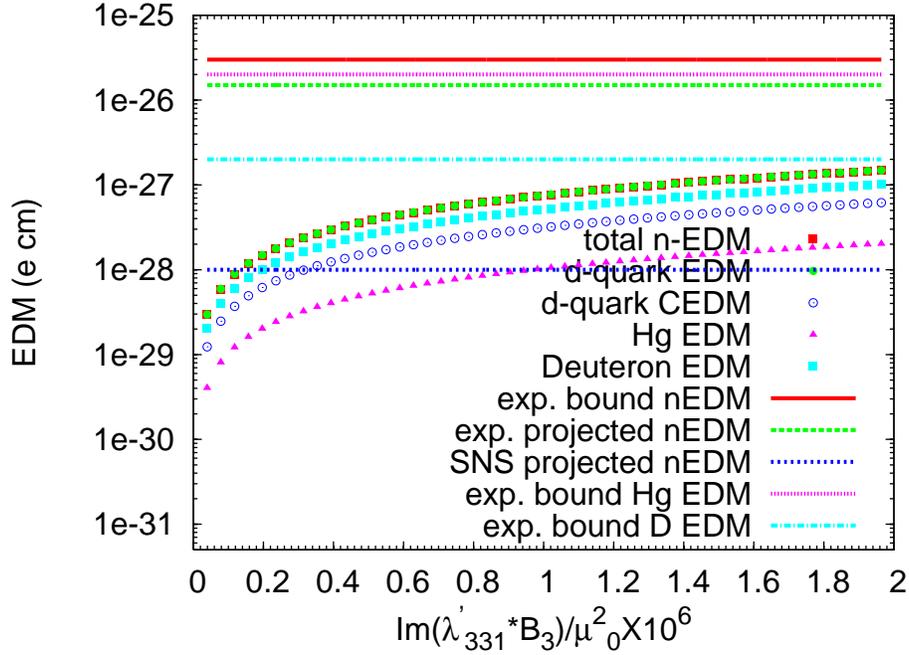}
\caption{(color online)
The neutron EDM (in units of e.cm) versus the combination
$\Im(B_3^{\ast}\l^{\prime}_{331})$ normalized by $\m_{\ssc 3}^2$.The
horizontal lines show present EDM bounds as well as improvements
possible in the future experiments.red solid line is current experimental
bound for neutron EDM \cite{nEDM-recent}, pink dotted line is current
experimental bound for mercury EDM \cite{Hg-bound}, green dashed line
is the goal expected to be reached by the current Grenoble neutron EDM
experiment
\cite{nEDM-goal}, cyan dash-dotted line is the expected measurement of
of the proposed deuteron EDM experiment \cite{dt-edm-proposal}, and
blue hashed line is the expected improvement at the proposed Los
Alamos neutron EDM experiment \cite{nEDM-SNS}. The red filled squares stand
for the total neutron EDM prediction, green filled circles are d-quark
EDM, blue open circles are d-quark CEDM, cyan filled squares are
predictions for deuteron EDM and magenta filled triangles are
predictions for mercury EDM, in the present framework described in the
previous section.}
\label{b3-331}
\end{figure}

In Fig.(\ref{b3-331}) we have plotted the neutron, deuteron, and
mercury EDMs, as well as d-quark EDM and CEDM, against the most
important combination ${\rm Im}(B_3^{\ast}\l^{\prime}_{331})$
normalized by $\m_{\ssc
  0}^2$. This combination has the largest impact owing to top Yukawa
and top mass dependence.  We have not plotted u-quark EDM as it is
about 5 orders of magnitude smaller than d-quark EDM (reflecting
up-top hierarchy). The description of various lines and symbols is
given in the caption for the figure. It is straight forward to
understand the features of the plot. From the
Eqs.(\ref{eq:edm-formula}) and (\ref{eq:cedm-formula}, given that
loop functions $F_3 \approx F_4$ for the parameters considered here,
one can see that $d^{\!\ssc E}_d/d^{\!\ssc C}_d \approx 5/(2\, g_s)
\approx 2.5$. This can be roughly observed in the plot. The plot
also shows that total neutron EDM is same as the d-quark EDM
contribution. To understand this, observe that u-quark EDM is
negligible and hence the total neutron EDM is sum of d-quark EDM and
CEDM. CEDM suffers not only a factor of 2.5 suppression as discussed
above, but also has additional $e/4\pi$ suppression, which more than
compensates QCD enhancement factor of 3.4. For d-quark EDM
contribution, QCD suppression factor is compensated by the
Clebsch-Gordan coefficient factor of $4/3$ and the total neutron EDM
works out to be roughly same as d-quark EDM. In the similar way one
can see from the formulas for deuteron and mercury EDMs that $d_{Hg}
\approx d^{\!\ssc C}_d/3.14$ and $d_{\mathrm{deuteron}} = 1.6\times
d^{\!\ssc C}_d$. All these relations are born out in the plot.

Coming to bounds, from Fig.\ref{b3-331}, we see that the null result
at the proposed Los Alamos EDM experiment can lead to a particularly
stringent constraint on ${\rm
Im}(B_3^{\ast}\l^{\prime}_{331})/\m_{\ssc
  0}^2 < 1.4 \times 10^{-7}$. For the same combination the bound we
obtain from the sensitivity of the proposed deuteron EDM experiment is
$4.0\times 10^{-6}$ and from the present best limits on neutron EDM
and Hg EDM experiment, $4.0\times 10^{5}$ and $1.9\times 10^{-4}$
respectively. This gives us an idea about the unprecedented
sensitivity of the proposed Los Alamos neutron EDM experiment and the
deuteron EDM experiment.

\begin{figure}[htbp]
  \centering
\includegraphics[scale=1]{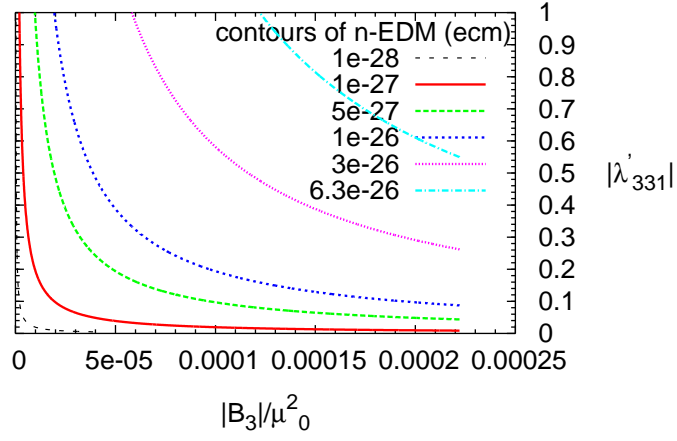}
  \caption{A contour plot for various values of neutron
    EDM in e.cm in the plane of real $\l^{\prime}_{331}$ and $B_3$.
    (with a relative phase of $\pi /4$).  Cyan dash-dotted line is the
    previous best experimental bound, pink dotted line is the current
    best bound, blue hashed line is the expected reach of Grenoble
    experiment, black dashed line is the Los Alamos projected
    sensitivity. One can see a sizable improvement in the regions
    ruled out in parameter space with improvement in sensitivity.}
  \label{fig:contour_n}
\end{figure}

\begin{figure}[htbp]
  \centering
\includegraphics[scale=1]{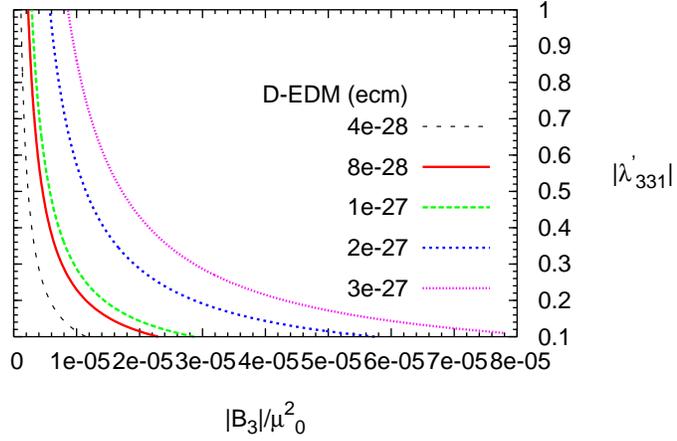}
  \caption{A contour plot for various values of deuteron
    EDM in e.cm in the plane of real $\l^{\prime}_{331}$ and $B_3$
    (with a relative phase of $\pi /4$). Region between pink dotted
    line and green dashed line is the expected sensitivity of the
    proposed deuteron experiment. Black dashed line shows a possible
    order of magnitude improvement in this. Again one can see that any
    null result can rule out huge regions in parameter space.}
  \label{fig:contour_dt}
\end{figure}

Fig.\ref{fig:contour_n} shows the contours of neutron EDM in the plane
of magnitudes for the couplings $\l^{\prime}_{331}$ and $B_3$ with the
relative phase fixed at $\pi/4$. The contour for present experimental
bound is shown in pink dotted line. The plot shows that a sizable
region of the parameter space are ruled out. Successive contours show
smaller values of neutron EDM with the smallest being $10^{-28}$ $e$
cm, the projected sensitivity of Los Alamos experiment.
Fig.\ref{fig:contour_dt} shows the contours of deuteron EDM in the
plane of magnitudes for the couplings $\l^{\prime}_{331}$ and $B_3$
with the relative phase fixed at $\pi/4$. Region between pink dotted
line and green dashed line is the expected sensitivity of the proposed
deuteron experiment. Black dashed line shows a possible order of
magnitude improvement in this. One can see that any null result
can rule out huge regions in parameter space.

\begin{figure}[htbp]
  \centering
  \includegraphics{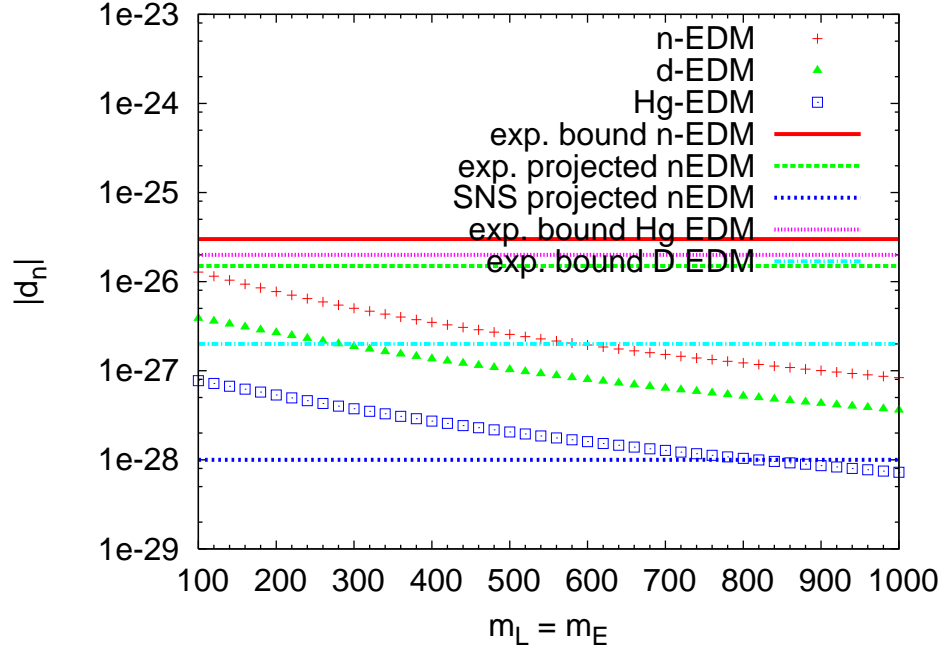}
  \caption{Various EDMs vs. the slepton mass parameter
    ${\tilde m}_{L} = {\tilde m}_{E}$ (with $|B_3| = 200
    ~\mbox{GeV}^2$, $|\l^{\prime}_{331}| = .05$ and relative phase of
    $\pi/4$).  Horizontal lines are the experimental bounds. }
  \label{fig:m3}
\end{figure}

\begin{figure}[htbp]
  \centering
  \includegraphics{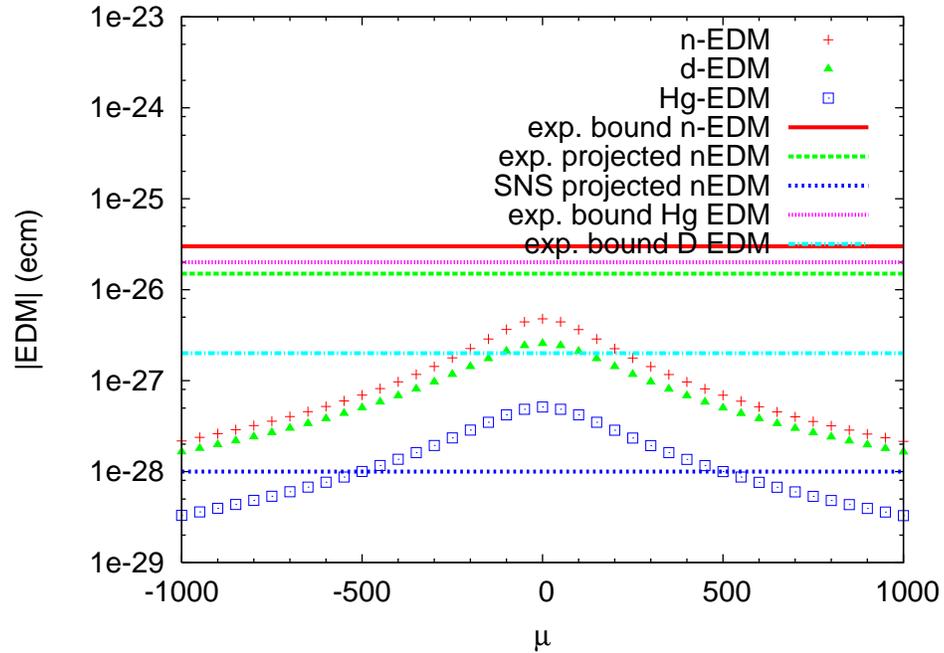}
  \caption{Various EDMs vs. $\m_{\ssc 0}$ parameter with
$|B_3| = 200 ~\mbox{GeV}^2$, $|\l^{\prime}_{331}| = .05$ and relative
phase $\pi/4$. Horizontal lines are the experimental bounds.}
  \label{fig:mu}
\end{figure}

So far we have kept certain parameters like the phase of the RPV
combination, the $\mu_{\ssc 0}$ parameter and the slepton mass
spectrum fixed.  To get a better understanding of the allowed regions
in the overall parameter space, let us focus on variations of the
parameters one at a time.  In Fig.(\ref{fig:m3}) we have plotted
neutron EDM versus the slepton mass parameter ${\tilde m}_{L} =
{\tilde m}_{E}$ (with $|B_3| = 200 ~\mbox{GeV}^2$,
$|\l^{\prime}_{331}| = .05$ and relative phase of $\pi/4$).  As
expected the neutron EDM falls with the increasing slepton mass.  More
careful checking reveals that the result is sensitive only to one
slepton mass parameter, the mass of the third $L$-handed slepton here.
It is also easy to understand from our analytical formulas that the
dominating contributions among the various scalar mass eigenstates for
the case of $B_i^{\ast}\l^{\prime}_{ijk}$ come from the $i$-th
$L$-handed slepton and the Higgs. Various horizontal lines are the
experimental bounds. It is interesting to note that even if slepton
mass is as high as 1 TeV, it still gives neutron EDM contribution that
is an order of magnitude larger than the Los Alamos experiment
sensitivity.  In Fig.(\ref{fig:mu}) we have shown the variation of
neutron EDM with the $\m_{\ssc 0}$ parameter.  Although parameter
$\m_{\ssc 0}$ does not directly figure in the EDM formula, its
influence is felt through the Higgs spectrum. Larger $\m_{\ssc 0}$
leads to heavier Higgs spectrum which suppresses the EDM contribution.
Again, it is interesting to see that $\m{\ssc 0} = 1$TeV, still gives
neutron EDM contribution that is larger than the sensitivity of Los
Alamos experiment.  The large value of neutron EDM (much larger than
the reach of Los Alamos experiment) even for slepton mass and $\m_{\ssc 0}$
parameter at $1$ TeV is not only encouraging but also
indicates that if a non-zero EDM is measured, one should exercise
caution in the interpretation in terms of any new physics model.
Our analytical formulas show that there is no
strong dependence on $\tan\! \b$ (we have checked this numerically as
well) and hence we have kept $\tan\! \beta =3$ fixed in all the plots.
 Relative insensitivity to the value of $\tan \!\beta$ here can be contrasted to
certain interesting predictions like Branching fraction for $B_s \to
\m^+ \m^-$ in MSSM which is boosted by three orders of magnitude in
large $\tan\! \beta$ region. Any constraints on $\tan \!\beta$
following non-observations of $B_s \to \m^+ \m^-$ would still leave
our predictions unaffected. The fact that our predictions largely
depend on hitherto unconstrained combination $B_i^*
\lambda^{\prime}_{ij1}$ and already known top Yukawa and top mass,
makes it all the more interesting.

\begin{figure}[htbp]
  \centering
\parbox{6.8in}{
\includegraphics[width=15cm]{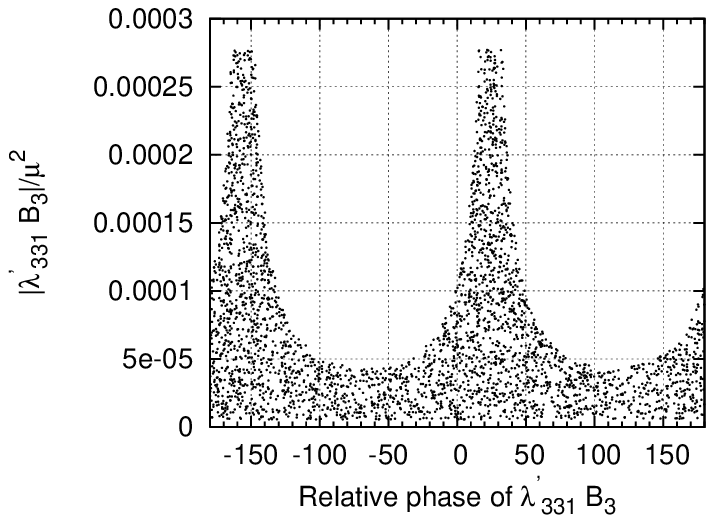}
\vskip 5mm
\includegraphics[width=15cm]{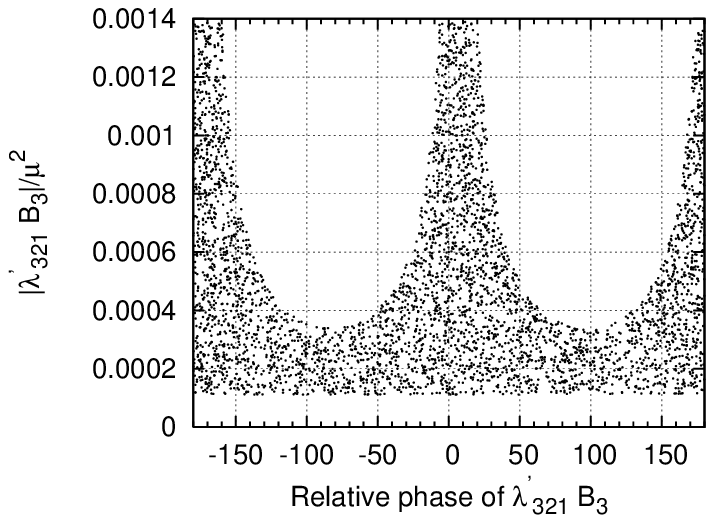}
\caption{
Dotted region is the parameter space
allowed by neutron EDM upper bound in the plane of
relative phase of $\l^{\prime}_{331}B^{\ast}_3$ and the
$|\l^{\prime}_{331}B^{\ast}_3|$ (top) and
relative phase of $\l^{\prime}_{321}B^{\ast}_3$ and the
$|\l^{\prime}_{321}B^{\ast}_3|$ (bottom).}
  \label{fig:scatter_n_331}}
\end{figure}

\begin{figure}[htbp]
  \centering
\parbox{6.8in}{
\includegraphics[width=15cm]{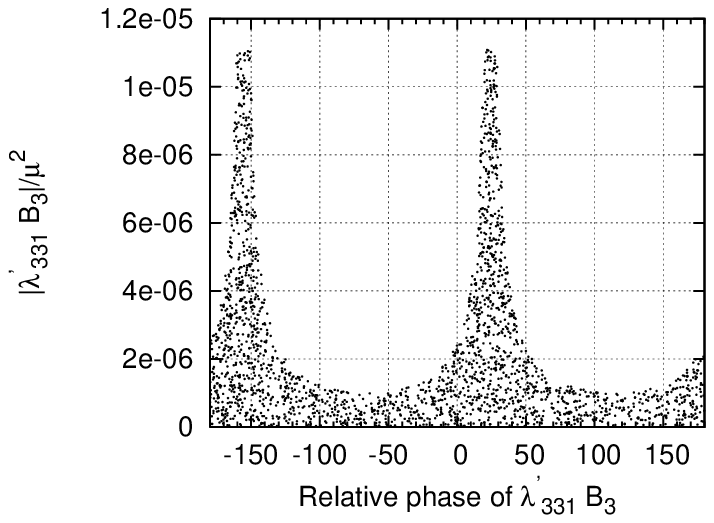}
\vskip 5mm
\includegraphics[width=15cm]{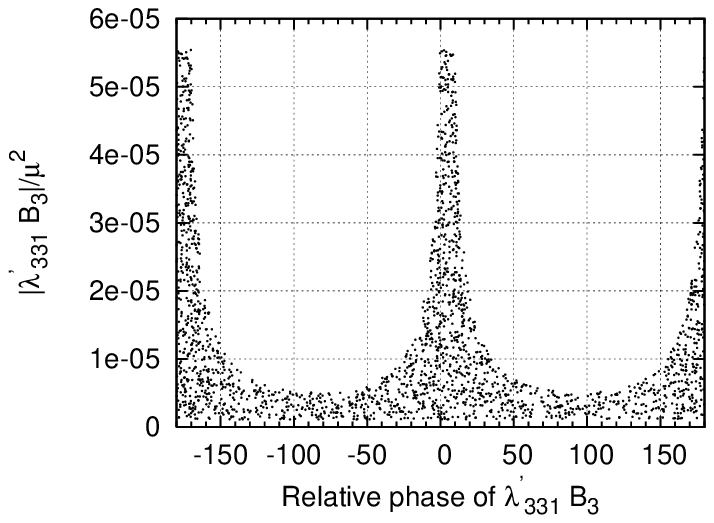}
\caption{
Dotted region is the parameter space
allowed by proposed deuteron EDM upper bound in the plane of
relative phase of $\l^{\prime}_{331}B^{\ast}_3$ and the
$|\l^{\prime}_{331}B^{\ast}_3|$ (top) and
relative phase of $\l^{\prime}_{321}B^{\ast}_3$ and the
$|\l^{\prime}_{321}B^{\ast}_3|$ (bottom).}
  \label{fig:scatter_dt_331}}
\end{figure}

To go beyond the illustrative case of the parameter combination
$B_i^{\ast}\l^{\prime}_{i31}$ , we list in table \ref{bounds} the
bounds due to the neutron EDM from the current Grenoble results (in
the column under I), due to plausible null result at Los Alamos
experiment (in the column under II), as well as due to any null
results of the future deuteron EDM experiment (in the column under
III), on the nine combinations $|B_i^{\ast} \l^{\prime}_{ij1}|$
normalized by $(100~\mbox{GeV})^2$. It is interesting to note the
difference in bounds for case (a) and (b) for coupling combination
$|B_i^{\ast} \l^{\prime}_{i31}|$ and
$|B_i^{\ast}\l^{\prime}_{i21}|$. The inputs for Case (b) are
identical to case (a) except for RPV phase being zero for case (b).
Case (b) thus relies solely on CKM phase. Interestingly the bound
for $|B_i^{\ast}\l^{\prime}_{i31}|$ changes very marginally from
case (a) to (b) whereas the bound for $|B_i^{\ast}
\l^{\prime}_{i21}|$ weakens by about an order of magnitude. To
understand this difference in behavior of
$|B_i^{\ast}\l^{\prime}_{i31}|$ and $|B_i^{\ast}\l^{\prime}_{i21}|$,
with and without a complex phase in the RPV couplings, we have
plotted in Fig. \ref{fig:scatter_n_331} the allowed region in the
plane of relative phase of $B^{\ast}_3 \l^{\prime}_{331}$ and the
$|B^{\ast}_3\l^{\prime}_{331}|$ (left) and relative phase of
$B^{\ast}_3 \l^{\prime}_{321}$ and the $|B^{\ast}_3
\l^{\prime}_{321}|$ (right). One can see that the bound for
$|B^{\ast}_3 \l^{\prime}_{331}|$ is about the same for relative
phase in $B^{\ast}_3 \l^{\prime}_{331}$ of 0 or $\pi/4$ but the
bound for $|B^{\ast}_3 \l^{\prime}_{321}|$ strengthens by about an
order of magnitude as the phase in the RPV coupling increases from 0
to $\pi/4$ suggesting a collaborative effect between the CKM phase
and the RPV phase.  Fig.\ref{fig:scatter_dt_331} is similar to
Fig.\ref{fig:scatter_n_331} except that it is now for deuteron EDM.
We see that a similar pattern here also because of the similar
qualitative dependence on down-quark EDM, implying similar
interference pattern between CKM phase and RPV phase induced
contributions. For the case of $B_i^{\ast} \l^{\prime}_{i11} $ there
is no CKM phase involved.  In the table we have fixed the sign of
$\m_{\ssc 0}$ parameter to be negative. In the Fig.\ref{fig:mu} it
is seen that absolute value of neutron EDM is symmetric with respect
to sign of $\m_{\ssc 0}$ parameter and hence positive $\m_{\ssc 0}$
should give identical bounds.  The variation of bounds in
table~\ref{bounds} with changes in parameters $\m_{\ssc 0}$ and the
slepton and Higgs mass very much follows the pattern found in plots
discussed earlier.  \vspace{0.5cm} {\small
\begin{table}
  \caption{Here we list the normalized upper bounds for several
    combinations of bilinear and trilinear RPV parameters, with some
    variation in the input parameters. The bounds essentially depend on the
    values of parameters like ${\tilde m}_{\ssc L},\m_{\ssc 0}$ and
    $m_{{\ssc H}_d}$ ($m_{{\ssc H}_u}$ and $B_{0}$ being determined from EW
    symmetry breaking condition). $\tan\! \b$ has been
    kept fixed at 3 as EDM has a very mild dependence on $\tan\! \b$. Note that
    all these bounds include CEDM contributions. Columns headings
    I,II,III refer to bounds from current neutron EDM experiment,
    from the possible null results at the Los Alamos neutron EDM
    experiment, and the possible null results at the deuteron EDM experiment respectively.
    As is clearly seen, any null results at Los Alamos experiment can
    lead to bounds that are over two orders of magnitude larger than the current bounds }
  \begin{tabular}{lcccc||c|c|c||c|c|c||c|c|c}
    \hline\hline
    \multicolumn{5}{c}{{\bf parameter values}} & \multicolumn{9}{c}{{\bf Normalized bounds}}\\\hline
    &$\m_{\ssc 0}$ & ${\tilde m}_{\ssc L}$ & $m_{{\ssc H}_d}$ &
    RPV  & \multicolumn{3}{c}{$\f{\Im(B_i^{\ast}\cdot \l^{\prime}_{i31})}{(100~\mbox{GeV})^2}$}  &
    \multicolumn{3}{c}{$\f{\Im(B_i^{\ast}\cdot \l^{\prime}_{i21})}{(100~\mbox{GeV})^2}$} &
    \multicolumn{3}{c}{$\f{\Im(B_i^{\ast}\cdot\l^{\prime}_{i11})}{(100~\mbox{GeV})^2})$}\\\cline{6-14}
    &(GeV)&(GeV)&(GeV)&phase&I&II&III&I&II&III&I&II&III\\\hline
    (a)&-100  & 100  & 100  & $\pi/4$
    & $1.5 \cdot 10^{-4}$&$ 5.0\cdot 10^{-7}$& $1.5\cdot10^{-5}$
    & $4.3\cdot 10^{-4}$&$1.3\cdot 10^{-6}$& $2.5\cdot 10^{-5}$
    &$ 1.7\cdot 10^{-3}$&$5.2\cdot 10^{-6}$&$1.1\cdot 10^{-4}$\\
    (b)\footnote{For this case the constraints correspond to the real part of
      RPV combination as the relative phase is zero.}
    &-100  & 100  & 100  & 0
    &$ 8.0\cdot 10^{-5}$&$2.5\cdot 10^{-7}$&$2.0\cdot 10^{-5}
    $&$2.0\cdot 10^{ -3}$&$6.5\cdot 10^{ -6}$&$5.1\cdot 10^{ -4}
    $&nil&nil&nil\\
    (c)&-400  & 100  & 100  & $\pi/4$
    &$5.5\cdot 10^{-4}$&$2.6\cdot 10^{-6}$&$5.2\cdot 10^{-5}
    $&$2.7\cdot 10^{-4}$&$9.2\cdot 10^{-6}$&$1.6\cdot 10^{-4}
    $&$1.0\cdot 10^{-2}$&$3.5\cdot 10^{-5}$&$6.6\cdot 10^{-4}$\\
    (d)&-800  & 100  & 100  & $\pi/4$
    &$1.7\cdot 10^{-3}$&$5.5\cdot 10^{-5}$&$1.5\cdot 10^{-4}
    $&$9.0\cdot 10^{-3}$&$3.0\cdot 10^{-5}$&$5.4\cdot 10^{-4}
    $&$3.2\cdot 10^{-2}$&$1.1\cdot 10^{-4}$&$2.1\cdot 10^{-3}$\\
    (e)&-100  & 400  & 100  & $\pi/4$
    &$6.3\cdot 10^{-4}$&$2.1\cdot 10^{-6}$&$5.6\cdot 10^{-5}
    $&$4.2\cdot 10^{-3}$&$1.5\cdot 10^{-5}$&$2.6\cdot 10^{-4}
    $&$1.5\cdot 10^{-2}$&$5.2\cdot 10^{-5}$&$1.1\cdot 10^{-3}$\\
    (f) &-100  & 800  & 100  & $\pi/4$
    &$1.9\cdot 10^{-3}$&$6.4\cdot 10^{-6}$&$1.6\cdot 10^{-4}
    $&$1.5\cdot 10^{-2}$&$5.2\cdot 10^{-5}$&$9.2\cdot 10^{-4}
    $&$5.1\cdot 10^{ -2}$&$1.8\cdot 10^{ -4}$&$3.6\cdot 10^{ -3}$\\
    (g)&-100  & 100  & 300  & $\pi/4$
    &$3.8\cdot 10^{ -4}$&$1.2\cdot 10^{ -6}$&$3.6\cdot 10^{ -5
    }$&$1.7\cdot 10^{ -3}$&$6.0\cdot 10^{ -6}$&$1.0\cdot 10^{ -4
    }$&$6.6\cdot 10^{ -3}$&$2.3\cdot 10^{ -5}$&$4.2\cdot 10^{ -4}$\\
    (h)&-100  & 100  & 600  & $\pi/4$
    &$1.0\cdot 10^{ -3}$&$3.5\cdot 10^{ -6}$&$9.2\cdot 10^{ -5
    }$&$5.8\cdot 10^{ -3}$&$2.0\cdot 10^{ -5}$&$3.5\cdot 10^{ -4
    }$&$2.1\cdot 10^{ -2}$&$7.6\cdot 10^{ -5}$&$1.4\cdot 10^{ -3}$\\
    \hline\hline
  \end{tabular}
  \label{bounds}
\end{table}}

Before we conclude, we would like to briefly comment on two things. In the
table~\ref{bounds}, we have only mentioned the bounds for the nine
combinations $| B_i^{\ast} \l^{\prime}_{ij1}|$, whereas earlier
in the text we did mention that in principle all the twenty-seven
$\l^{\prime}_{ijk}$ together with bilinear couplings do contribute to
the neutron EDM. The couplings other than $\l^{\prime}_{ij1}$ contribute
to $u$ quark EDM and all the contributions are proportional to
up-Yukawa (in contrast to the presence of a contribution proportional to top-Yukawa
for the $d$ quark EDM). Strength of the corresponding contributions is substantially
weaker (typically by 5 to 6 orders of magnitude) and hence do not lead to
meaningful bounds. The other thing is about the possible $\mu_i^{\ast}
\cdot \l^{\prime}_{ijk}$ contributions. It can be seen in
Eq.(\ref{mui}) that these are second order in perturbation and also
accompanied by lepton mass $\m_i$.
Thus, a typical $\m_i^{\ast} \l^{\prime }_{ijk}$
contribution is substantially weaker than the corresponding
$B_i^{\ast}  \l^{\prime }_{ijk}$ contribution. For instance,
with the similar inputs for other SUSY parameters as described above,
if one takes $\mu_{3} = 10^{-3}$ GeV (dictated by requirement of sub-ev
neutrino masses), $\l^{\prime}_{331} = .05$ and a relative phase of
$\pi /4$, one obtains neutron EDM of $6.0 \cdot 10^{-32}$ which is
about six orders of magnitude smaller than the present experimental
bound on neutron EDM. If one goes by the upper bound on the mass of
$\n_{\tau}$ of 18.2 MeV \cite{nu_mass}, $\mu_{\ssc 3}$ could be as large as 7
GeV for $\tan\! \b = 2$ and sparticle mass $\sim 300$ GeV
\cite{ru2}. For $\m_{\ssc 3}= 1$
GeV we obtain neutron EDM of $6.1 \cdot 10^{-29}$, still about three
orders of magnitude smaller than the experimental bound.
These numbers can be easily compared with the $\m^{\ast}_i
\l^{\prime}_{i11}$ contributions to the $d$ quark EDM through
chargino loop in Table~1 of ref.\cite{nedm}. There the corresponding
contribution (with $\m_{\ssc 3} = 1$ GeV) is much weaker (of order
$10^{-32}$) as it lacks the top Yukawa and top mass enhancements. In
the same table one finds that corresponding contribution to gluino
loop is much stronger (of order $10^{-25}$) due to proportionality to
gluino mass and strong coupling constant. In the light of above reasons
one can appreciate that the contributions due to soft parameter $B_i$
are far more dominating in the present scenario of quark loop
contributions.

\section{Conclusions}
We have made a systematic study of the influence of the combination
of bilinear and trilinear RPV couplings on the neutron, deuteron,
and mercury EDMs, including the contributions from chromo-electric
dipole form factor.  Such combinations are interesting because this
is the only way RPV parameters contribute at one-loop level to EDM.
The fact that the form factors are proportional to top Yukawa and
top mass with the hitherto unconstrained combination of RPV
parameters, makes them all the more interesting.
Such class of diagrams have a quark as the fermion running inside
the loop.  EDMs of neutron, deuteron, and mercury are ultimately
expressed in terms of quark EDMs and CEDMs.  In our analytical
expressions obtained based on perturbative diagonalization of the
scalar mass-squared matrices, we demonstrated that charged-slepton
Higgs mixing loop contribution to $d$ quark EDM far dominates the
other contributions due to a diagram with the top-quark in the loop.
In our numerical exercise we have obtained robust bounds on the
combinations $\f{{\rm Im}(B_i^{\ast}\cdot
\l^{\prime}_{ij1})}{(100~\mbox{GeV})^2} $ for $i,j=1,2,3$ that have
not been reported before. The bounds are reported for the current
best limits on neutron EDM as well as projected for any null results
in the future improvements at Grenoble and Los Alamos experiments as
well as deuteron EDM experiment. It turns out that measurements of
mercury EDM are not as strongly constraining as current neutron EDM
limits, whereas any null result at deuteron EDM and/or Los Alamos
neutron EDM experiment can lead to much stronger constraints on RPV
parameter space (for instance${\rm Im}
(B_3^{\ast}\l^{\prime}_{331})/\m_{\ssc 0}^2 < 1.4 \times 10^{-7}$).
Even if the RPV couplings are real, they could still contribute to
quark EDMs via CKM phase. For some cases CKM phase induced
contribution is as strong as that due to an explicit complex phase
in the RPV couplings. We find that even if slepton mass or
$\mu_{\ssc 0}$ are as high as 1 TeV, it could still lead to bounds
that are well within the reach of Los Alamos neutron EDM experiment.
There also exist contributions involving $\m_i^{\ast}
\l^{\prime}_{ijk}$. However these are higher order effects which are
further suppressed by proportionality to charged lepton mass. Since
$\m_i$ are expected to be very small (of order $10^{-3}$ GeV) for
sub-eV neutrino masses, such contributions are highly suppressed.
Our results presented here make available a new set of interesting
bounds on combinations of RPV parameters.

One further thing worth some attention here is the question of to what
extent our choice of model formulas for the hadronic EDMs from quark dipoles,
as discussed and justified above, affect our major conclusions 
\footnote{The question is prompted by a journal referee, to whom we
express our gratitute.}. One can think about presenting full numerical results
and comparison for the cases of the various different model formulas. 
However, we find that neither feasible nor desirable. We did do some numerical
checking, but consider a brief summary here as the only appropriate thing
to present. Two special features of the RPV model dictate the answer. The
hadron EDMs are to be determined from the EDMs and CEDMs of the $u$, $d$,
and $s$ quark. For the RPV model, the resulted dipoles for $u$-quark are much 
smaller than of the $d$-quark. For the $s$-quark, however, dominant contributions
involve similar diagrams with trilinear RPV couplings of different familiy 
indices, typically replacing a $1$ for the $d$ by a $2$ for the $s$. The latter
complicaton makes it difficult to fully address the impact within the scope of
the present study. However, one can at least see that for the RPV parameter 
combinations playing a dominant role in generating $d$ dipoles, our main focus 
here, their contributions to the $s$ dipoles are going to be unimportant.
Neglecting the the $s$ and $u$ dipoles, models formulas from QCD sum rule and 
the valence quark model give roughly $d_n\simeq 0.92 \, d_d^{\!\ssc E}$ and
$d_n\simeq 0.96 \, d_d^{\!\ssc E}$, respectively. The case for comparison
with the chiral lagrangian approach is more complicated. But the result
agrees within a factor of three with the two numbers. The deuteron EDM
numbers from the two approaches agree within error bars. And the Mercury
EDM constraint is always negligible when compared to that of neutron and
Deuteron, even when $s$ dipoles are to be considered. Hence, our results here 
are not much affected by the hadron EDMs model formulas chosen.


\newpage

\end{document}